\documentclass[aps,pra,nofootinbib]{revtex4}
\usepackage{amsmath,amssymb}
\usepackage{graphicx}
\usepackage{pst-plot}
\usepackage{mathrsfs}
\usepackage{appendix}
\usepackage{mathtools}
\usepackage{pst-func}

\usepackage{chemarrow}

\usepackage{rotating}

\begin{document}

\title{Restoring Local Causality and Objective Reality to the Entangled Photons}

\author{Joy Christian}

\email{joy.christian@wolfson.ox.ac.uk}

\affiliation{Department of Physics, University of Oxford, Parks Road, Oxford OX1 3PU, United Kingdom}

\begin{abstract}
Unlike our basic theories of space and time, quantum mechanics is not a locally causal theory. Moreover,
it is widely believed that any hopes of restoring local causality within a realistic theory have been
undermined by Bell's theorem and its supporting experiments. By contrast, we provide a strictly local,
deterministic, and realistic explanation for the correlations observed in two such supporting experiments,
performed independently at Orsay and Innsbruck. To this end, a pair of local variables is constructed to
simulate detections of photon polarizations at various angles, chosen freely by Alice and Bob. These generate
purely random outcomes, ${{\mathscr A}=\pm\,1}$ and ${{\mathscr B}=\pm\,1}$, occurring within a
parallelized 3-sphere.
These outcomes do not depend on the contexts of measurements, but are determined entirely by the randomly chosen
initial orientation of the 3-sphere. When they are compared, however, the correlation between
them turn out to be exactly equal to
${-\cos2(\alpha - \beta)}$, with the corresponding CHSH inequality
violated for the polarization angles ${\alpha}$, ${\alpha'}$, ${\beta}$, and ${\beta'}$ in precisely the manner
predicted by quantum mechanics. The key ingredient in our explanation is the topology of the 3-sphere, which remains
closed under multiplication, thus preserving the locality condition of Bell. It allows us to model the physical space
as a 3-sphere,
and reveals that the illusion of quantum nonlocality in the present case stems from a twist in the
Hopf fibration of the 3-sphere.
\end{abstract}

\maketitle

\parskip 5pt

{\baselineskip 12.53pt

One of the first steps we often take towards measuring a physical quantity is to set up a Cartesian
coordinate system ${\{x,y,z\}}$ in the Euclidean space ${{\mathbb E}_3}$. This amounts to
modeling the Euclidean space as a 3-fold product of the real line, ${{\rm I\!R}^3}$. This
procedure has become so familiar to us that in practice we often identify ${{\mathbb E}_3}$
with its Cartesian model, and simply think of ${{\rm I\!R}^3}$ as {\it the} Euclidean
space. As we shall see, however, this seemingly innocuous act of convenience comes with a
heavy price: It is largely responsible for the illusion of quantum nonlocality \cite{EPR}\cite{Bell-1964}.
In what follows we shall demonstrate that,
once a correct coordinate-free geometric model of the physical space is used, the correlations observed in
the EPR-inspired experiments involving photon pairs \cite{Peres-1993}\cite{Aspect-666}\cite{Weihs-666}, namely
\begin{align}
{\mathscr A}(\alpha) = \pm\,1,\;{\mathscr B}(\beta) = \pm\,1,\;\; \notag \\
{\cal E}(\alpha) = 0,\;\;\; {\cal E}(\beta) = 0,\;\;\;\;\;\, \notag \\
{\cal E}(\alpha,\,\beta) \,=\,-\cos2(\alpha - \beta),\;\,\label{q}
\end{align}
are easily understood, in a strictly local-realistic terms. Here ${{\cal E}(\alpha,\,\beta)}$
represents the expected value of simultaneously observing remote measurement results
${{\mathscr A}(\alpha)}$ and ${{\mathscr B}(\beta)}$ along the polarization angles ${\alpha}$ and ${\beta}$, respectively.

Euclid himself of course did not think of ${{\mathbb E}_3}$ in terms of triples of real numbers.
He defined its representation axiomatically, entirely in terms of primitive geometric objects
such as points and lines, together with a list of their properties,
from which his theorems of geometry follow. Today we know, however, that it is quite problematic
to give a suitable definition of Euclidean space in the spirit of Euclid, and hence in physics
we instinctively identify ${{\mathbb E}_3}$ with ${{\rm I\!R}^3}$ whenever possible. But there is no
natural, geometrically-determined way to identify the two spaces without introducing an {\it unphysical}${\,}$
notion of arbitrarily distinguished coordinate system \cite{Eberlein}. This difficulty is clearly relevant in the
study of Bell's theorem \cite{Bell-1964}, for time and again we have learned that surreptitious introduction
of unphysical ideas in physics could lead to distorted views of the physical reality. A coordinate-free
representation of the Euclidean space is undoubtedly preferable, if what is at stake is the very
nature of the physical reality.

Fortunately, precisely such a representation of ${{\mathbb E}_3}$, with a rich algebraic structure, was provided
by Grassmann in 1844 \cite{Clifford}. As in Euclid's geometry, the basic elements of this powerful structure
are not coordinate systems, but points, lines, planes, and volumes, {\it all treated on equal footing}. Today one
begins this framework by postulating a unit volume element (or a trivector) in ${{\mathbb E}_3}$, defined by
\begin{equation}
I={{\bf e}_x}\wedge\,{{\bf e}_y}\wedge\,{{\bf e}_z}\,,\label{2}
\end{equation}
with ${\{{\bf e}_x,\,{\bf e}_y,\,{\bf e}_z\}}$ being a set of orthonormal vectors, ``${\wedge}$''
the outer product of Grassmann, and ${I^2= -1}$ \cite{Clifford}. Each vector ${{\bf e}_j}$ is then a solution of
the equation ${I\wedge{\bf e}_j\,=\,0}$, and every pair of them respects the fundamental geometric product
\begin{equation}
{\bf e}_j\,{\bf e}_k\,=\,{\bf e}_j\cdot{\bf e}_k+\,{\bf e}_j\wedge\,{\bf e}_k\,,\label{b-algebra}
\end{equation}
where ``${\,\cdot\,}$'' represents the inner product,
and ${{\bf e}_j\wedge\,{\bf e}_k}$ are unit bivectors (which also square to ${-1}$) with counterclockwise sense for the cyclicly
permuted indices (${j,\,k=x,\,y,\;{\rm or}\;z}$). The resulting {\it real} geometric structure is a
linear vector space spanned by the basis
\begin{equation}
\left\{1,\,{\bf e}_x,\,{\bf e}_y,\,{\bf e}_z,\,{\bf e}_x\wedge{\bf e}_y,\,
{\bf e}_y\wedge{\bf e}_z,\,{\bf e}_z\wedge{\bf e}_x,\,
{\bf e}_x\wedge{\bf e}_y\wedge{\bf e}_z\right\}\!,
\end{equation}
which encodes a graded linear algebra of dimensions eight. This algebra intrinsically characterizes the space ${{\mathbb E}_3}$.

Our interest, however, lies in a certain subalgebra of this algebra, the so-called even subalgebra
of dimensions four, defined by the bivector (or spinor) basis \cite{Clifford}\cite{Eberlein}:
\begin{equation}
\left\{1,\,{\bf e}_x\wedge{\bf e}_y,\,
{\bf e}_y\wedge{\bf e}_z,\,{\bf e}_z\wedge{\bf e}_x\right\}\!.\label{5}
\end{equation}
Crucially for our purposes, this subalgebra happens to remain closed under multiplication. Consequently, it can be used by
itself to model the physical space. In fact, it provides the most natural coordinate-free representation of the physical
space as a parallelized 3-sphere \cite{Eberlein}, which differs from ${{\rm I\!R}^3}$ only by a single point:
${S^3 = {\rm I\!R}^3\cup\{\infty\}}$. Both ${{\rm I\!R}^3}$ and ${S^3}$ are three-dimensional manifolds that are parallelized
by a vanishing Riemann curvature, but unlike in ${{\rm I\!R}^3}$ the torsion within ${S^3}$ is non-vanishing \cite{What-666}.
${S^3}$ is thus a one-point compactification of ${{\rm I\!R}^3}$ that remains as flat as ${{\rm I\!R}^3}$ \cite{Nakahara}.

The vectors and trivectors are no longer intrinsic to the above subalgebra, but belong to a dual space. Only the scalars and
bivectors---{\it treated on equal footing}---are taken to be intrinsic parts of the subalgebra. This can be seen more clearly
if we use the condition ${I\wedge{\bf e}_j=0\,}$ to rewrite the basis bivectors defined in Eq.${\,}$(\ref{5}) as
${I\cdot{\bf e}_z}$, ${I\cdot{\bf e}_x}$, and ${I\cdot{\bf e}_y}$. Their geometric product, analogous to the one in
Eq.${\,}$(\ref{b-algebra}), then leads to the defining equation of this subalgebra:
\begin{equation}
(I\cdot{\bf e}_j)\,(I\cdot{\bf e}_k)\,=\,-\;\delta_{jk}\,-\,\epsilon_{jkl}\;(I\cdot{\bf e}_l).\label{6}
\end{equation}
Evidently, despite the occurrences of trivectors and basis vectors, only the basis scalar and bivectors are involved in this
definition. Consequently, in what follows only scalars and bivectors (and their combinations) will have direct
physical significance---vectors and trivectors will merely facilitate computational ease, or ``hidden'' variables.

Given the bivector basis defined by Eq.${\,}$(\ref{5}), any generic bivector such as
${I\cdot{\bf a}}$ can be expanded in this basis as
\begin{equation}
I\cdot{\bf a}\,=\,
\{\,a_x\;{{\bf e}_y}\,\wedge\,{{\bf e}_z}
\,+\,a_y\;{{\bf e}_z}\,\wedge\,{{\bf e}_x}
\,+\,a_z\;{{\bf e}_x}\,\wedge\,{{\bf e}_y}\}.
\label{mu}
\end{equation}
In many ways bivectors---and also vectors and trivectors---behave just like ordinary numbers within this geometrical
framework. For this reason Grassmann referred to them as ``extensive magnitudes'', and treated them with the same respect
as he would treat ordinary numbers. And for the same reason Hestenes today refers to all such quantities as
``directed numbers'' \cite{Clifford}. They are simply numbers of higher grades and dimensions, with built-in directional attributes.

It is also worth noting that, although there is isomorphism between the vector subspace and the bivector
subspace, a bivector is an abstract entity of its own, with properties quite distinct from those of a vector \cite{Clifford}.
In fact it is a skeleton of only three properties: (1) a sense, quantified by a ${+}$ or ${-}$ sign, indicating whether
it represents a counterclockwise or clockwise rotation, (2) a magnitude,
which is equal to unity for all bivectors we will be considering, and (3) a direction, which can be specified by a dual vector
normal to its shape-independent plane. Thus, despite appearances, neither the trivector ${I}$ nor the vector
${\bf a}$ is an intrinsic part of the bivector ${I\cdot{\bf a}}$. Moreover, it is easy to verify that unit bivectors such as
${I\cdot{\bf b}}$ simply represent intrinsic points of a unit 2-sphere contained within the 3-sphere:
\begin{align}
||I\cdot{\bf b}\,||^2 &= (+\,I\cdot{\bf b})(-\,I\cdot{\bf b}) \notag \\
&= -\,I^2\,{\bf b}\,{\bf b} = {\bf b}\,{\bf b}
= {\bf b}\cdot{\bf b} =||{\bf b}||^2 = +\,1\,. \label{14-2}
\end{align}
More precisely, they represent the equatorial points of a parallelized 3-sphere modeling the physical space \cite{What-666}.

Given two such unit bivectors, say ${I\cdot{\bf a}}$ and ${I\cdot{\bf b}}$, the bivector subalgebra (\ref{6})
leads to the well known identity
\begin{equation}
(I\cdot{\bf a})(I\cdot{\bf b})\,=\,-\,{\bf a}\cdot{\bf b}\,-\,I\cdot({\bf a}\times{\bf b}),\label{i}
\end{equation}
provided we use the duality relation ${{\bf a} \wedge {\bf b}\,=\,I\cdot({\bf a}\times{\bf b})}$ defined between the vector
${{\bf a}\times{\bf b}}$ and the bivector ${{\bf a} \wedge {\bf b}}$.

The above identity provides a natural representation of points of a parallelized 3-sphere \cite{What-666}. The bivectors
${I\cdot{\bf a}}$ and ${I\cdot{\bf b}}$ appearing on its LHS represent the equatorial points of the 3-sphere, and the
{\it real} quaternion appearing on its RHS---which is a sum of a scalar and a bivector---represents a non-equatorial point
of the same sphere. An equator of a parallelized 3-sphere, however, which is a 2-sphere, does not remain closed under
multiplication. The 3-sphere itself, on the other hand, {\it does} remain closed under multiplication, thereby correctly encoding
the topology underlying the subalgebra (\ref{6}). Conversely, any arbitrary point ${P}$ of a parallelized 3-sphere can
always be factorized into any number of points: ${P=ABCD...}$ Needless to say, this is a highly nontrivial and powerful property
of the 3-sphere. To appreciate its non-triviality, consider a product of infinitely many points of a 3-sphere.
Such a product will simply be another point of the 3-sphere \cite{What-666}. By contrast, this will not be true in the case of
a 2-sphere even for just two points, as is\break evident from the above identity. The fact that both 3-sphere and its algebraic
representation (\ref{i}) remain closed under multiplication suggests that we should represent measurement results in the
present case by local maps of the form
\begin{equation}
{\mathscr A}({\bf a},\,\lambda): {\rm I\!R}^3\!\times\Lambda\longrightarrow S^2 \subset S^3, \label{map}
\end{equation}
where ${\Lambda}$ is a space of complete states ${\lambda}$, vector ${{\bf a}\in{\rm I\!R}^3}$ specifies
the context of measurement, and ${S^2}$ is an equatorial 2-sphere within a unit, parallelized 3-sphere.
Then not only the measurement results, but also their products would remain within the 3-sphere, thereby respecting
the locality (or factorizability) condition of Bell \cite{Bell-1964}. We shall see that a local-realistic interplay
between the points of such a 3-sphere and its equatorial 2-sphere is what is truly
responsible for the EPR correlations manifested in nature.

So far we have considered the bivector subalgebra (\ref{6})
with arbitrarily fixed basis, as in definition (\ref{5}). The convention
usually is to assume a right-handed set of basis bivectors, and so far we have followed this convention. The algebra itself,
however, does not fix the handedness of the basis. We could have equally well started out with a left-handed set of bivectors,
by letting ${-\,I}$ instead of ${+\,I}$ fix the basis. Equation (\ref{6}) would have then had the alternate form:
\begin{equation}
(-I\cdot{\bf e}_j)\,(-I\cdot{\bf e}_k)\,=\,-\;\delta_{jk}\,-\,\epsilon_{jkl}\;(-I\cdot{\bf e}_l).
\end{equation}
Comparing this equation with equation (\ref{6}) we see that there
remains a sign ambiguity in the definition of our subalgebra ({\it cf}.
Refs.${\,}$\cite{Eberlein} and \cite{Clifford}):
\begin{equation}
(I\cdot{\bf e}_j)\,(I\cdot{\bf e}_k)\,=\,-\;\delta_{jk}\,\pm\,\epsilon_{jkl}\;(I\cdot{\bf e}_l).\label{g}
\end{equation}
Consequently, following the time-honored mathematical practice of turning an ambiguity of sign into virtue, we define the
handedness of this entire subalgebra as our ``hidden variable.''
In other words, we specify the complete state of the photons we are about to study as ${\,{\boldsymbol\mu}=\pm\,I,\,}$
thereby defining the basis of our entire subalgebra by the equation
\begin{equation}
({\boldsymbol\mu}\cdot{\bf e}_j)\,({\boldsymbol\mu}\cdot{\bf e}_k)\,=
\,-\;\delta_{jk}\,-\,\epsilon_{jkl}\;({\boldsymbol\mu}\cdot{\bf e}_l).\label{14}
\end{equation}
The identity (\ref{i}) for the generic bivectors then becomes
\begin{equation}
(\,{\boldsymbol\mu}\cdot{\bf a})(\,{\boldsymbol\mu}\cdot{\bf b})\,
=\,-\,{\bf a}\cdot{\bf b}\,-\,{\boldsymbol\mu}\cdot({\bf a}\times{\bf b}),\label{bi-identity}
\end{equation}
along with {\it indefinite} duality relation ${{\bf a}\wedge{\bf b}={\boldsymbol\mu}\cdot({\bf a}\times{\bf b})}$. In
other words, the duality between the wedge product and cross product within our subalgebra alternates
between the right-hand and left-hand rules \cite{What-666}. Clearly, then, our complete state ${{\boldsymbol\mu}=\pm\,I}$
represents a far deeper hidden structure than the variables considered by Bell \cite{Bell-1964}. It un-fixes the orientation
of the entire physical space ${S^3}$, and turns it into a shared randomness between Alice and Bob.

We are now well equipped to take up the question of EPR correlations exhibited by the pair of entangled
photons in the actual experiments \cite{Aspect-666}\cite{Weihs-666}.
This question, of course, has been well scrutinized in the literature \cite{Peres-1993}.
We will restrict to the most basic aspects of the question, and follow its treatment given in Refs.${\,}$\cite{Peres-1993} and
\cite{Weihs-666}.
In a quantum mechanical description of the experiment involving photon pairs one usually assumes that the
system has been prepared in the singlet state
\begin{equation}
|\,\Psi_-\rangle\,=\,\frac{1}{\sqrt{2}\;}\Bigl\{|\,H\,\rangle_1\otimes|\,V\,\rangle_2\,
-\,|\,V\,\rangle_1\otimes|\,H\,\rangle_2\Bigr\}\,,\label{single}
\end{equation}
where ${|\,H\,\rangle}$ and ${|\,V\,\rangle}$ denote the horizontal and vertical polarization states of the photons
along the directions ${{\bf e}_x}$ and ${{\bf e}_y}$, respectively; and the subscripts 1 and 2 refer to the photons 1 and 2,
respectively. The photons are thus assumed to be propagating in the ${{\bf e}_z}$ direction. We could equally well
consider the polarization state ${|\,\Psi_+\rangle}$, but that would not add anything significant to our concerns here.
Both polarization states, ${|\,\Psi_+\rangle}$ and ${|\,\Psi_-\rangle}$, are invariant under rotations about the ${{\bf e}_z}$
axis, but the state ${|\,\Psi_+\rangle}$ is even under reflections, whereas the state ${|\,\Psi_-\rangle}$ is odd
under reflections.

In a typical experimental run Alice and Bob measure polarizations of the photons along two different directions in the plane
perpendicular to the ${{\bf e}_z}$ axis. Alice measures polarizations along the direction ${\bf a}$, which makes an angle
${\alpha}$ with the ${{\bf e}_x}$ axis, whereas Bob measures polarizations along the direction ${\bf b}$, which makes an
angle ${\beta}$ with the ${{\bf e}_x}$ axis. Individually, the binary results observed by Alice and Bob, namely
${A(\alpha) = \pm\,1}$ and ${B(\beta) = \pm\,1}$, are found to be completely random, with equal probabilities for
the outcomes ${+\,1}$ and ${-\,1}$.
When these results are compared, however, they are found to be strongly correlated,
in agreement with the quantum mechanical predictions we have summarized in Eq.${\,}$(\ref{q}).

Our goal now is to reproduce these quantum mechanical predictions {\it exactly}, within the geometrical model of
the physical space discussed above.
To this end, we have assumed that the complete state of the photons is given by ${{\boldsymbol\mu}=\pm\,I}$, where ${I}$
is the fundamental trivector defined in Eq.${\,}$(\ref{2}). The detections of photon polarizations observed by Alice and
Bob along their respective axes ${\bf a}$ and ${\bf b}$, with the bivector basis fixed by the trivector ${\boldsymbol\mu}$,
can then be represented {\it intrinsically} as points of the physical space ${S^3}$, by the following two local variables:
\begin{equation}
S^3\ni {\mathscr A}(\alpha,\,{\boldsymbol\mu})\,=\,(-\,I\cdot{\widetilde{\bf a}}\,)
\,(\,+\,{\boldsymbol\mu}\cdot{\widetilde{\bf a}}\,)\,=\,
\begin{cases}
+\,1\;\;\;\;\;{\rm if} &{\boldsymbol\mu}\,=\,+\,I \\
-\,1\;\;\;\;\;{\rm if} &{\boldsymbol\mu}\,=\,-\,I
\end{cases} \label{17-Joy}
\end{equation}
and
\begin{equation}
S^3\ni {\mathscr B}(\beta,\,{\boldsymbol\mu})\,=\,(+\,I\cdot{\widetilde{\bf b}}\,)
\,(\,+\,{\boldsymbol\mu}\cdot{\widetilde{\bf b}}\,)\,=\,
\begin{cases}
-\,1\;\;\;\;\;{\rm if} &{\boldsymbol\mu}\,=\,+\,I \\
+\,1\;\;\;\;\;{\rm if} &{\boldsymbol\mu}\,=\,-\,I\,,
\end{cases} \label{18-Joy}
\end{equation}
with equal probabilities for ${\boldsymbol\mu}$ being either ${+\,I}$ or ${-\,I}$,
and the rotating vectors ${\widetilde{\bf a}}$ and ${\widetilde{\bf b}}$ defined as
\begin{align}
&{\widetilde{\bf a}}= {\bf e}_x\,\cos2\alpha\,+\,{\bf e}_y\,\sin2\alpha\,, \notag \\
\text{and}\;\;\;
&{\widetilde{\bf b}}= {\bf e}_x\,\cos2\beta\,+\,{\bf e}_y\,\sin2\beta\,.\label{b-def}
\end{align}
Note that ${{\mathscr A}(\alpha,\,{\boldsymbol\mu})}$ and ${{\mathscr B}(\beta,\,{\boldsymbol\mu})}$,
in addition to being manifestly realistic, are strictly {\it local} variables\footnote{Needless
to say, ${{\mathscr A}({\alpha},\,{\boldsymbol\mu})}$ and ${{\mathscr B}({\beta},\,{\boldsymbol\mu})}$ are two
{\it different} functions of the random variable ${\boldsymbol\mu}$. Moreover, they are {\it statistically independent
events} occurring within a 3-sphere, with factorized joint probability
${P({\mathscr A}\;\text{and}\;{\mathscr B})=P({\mathscr A})\!\times\!P({\mathscr B})\leq\frac{1}{2}}$.
Therefore their product ${{\mathscr A}{\mathscr B}}$ is guaranteed to be equal to ${-1}$ only for the special case
${{\alpha}={\beta}}$. For all other ${\alpha}$ and ${\beta}$, ${{\mathscr A}{\mathscr B}}$ will
alternate between the values ${-1\;\text{and}\;+\!1}$.}.
In fact, they are not even contextual \cite{Contextual}.
Alice's measurement result, although refers to a freely chosen angle ${\alpha}$, depends only on the
initial state ${{\boldsymbol\mu}}$; and likewise, Bob's measurement result, although refers to a freely chosen angle
${\beta}$, depends only on the initial state ${{\boldsymbol\mu}\,}$. In other words, all possible measurement results at all
possible angles are completely determined by the initial orientation of the physical space specified
by ${{\boldsymbol\mu}}$, or equivalently by that of
the 3-sphere. Moreover, it is easy to check using the identity
(\ref{bi-identity}) that ${{\mathscr A}(\alpha,\,{\boldsymbol\mu})}$ and ${{\mathscr B}(\beta,\,{\boldsymbol\mu})}$
are purely binary numbers. For consider the product
\begin{align}
(+\,I\cdot{\widetilde{\bf a}}\,)(+\,I\cdot{\widetilde{\bf a}\,'})\,
&=\,-\,{\widetilde{\bf a}}\cdot{\widetilde{\bf a}\,'}\,-\,I\cdot(\,{\widetilde{\bf a}}\times{\widetilde{\bf a}\,'\,}) \notag \\
&=\,-\cos\theta_{{\widetilde{\bf a}}{\widetilde{\bf a}\,'}}\,-\,\left(\,I\cdot{\widetilde{\bf c}}
\,\right)\,\sin\theta_{{\widetilde{\bf a}}{\widetilde{\bf a}\,'}}\,, \label{quat}
\end{align}
where ${{\widetilde{\bf c}}={\widetilde{\bf a}}\times{\widetilde{\bf a}\,'}/|{\widetilde{\bf a}}\times{\widetilde{\bf a}\,'}|}$,
and ${\theta_{{\widetilde{\bf a}}{\widetilde{\bf a}\,'}}}$ is the angle between ${\widetilde{\bf a}}$ and ${\widetilde{\bf a}\,'}$.
The RHS of this identity is a quaternion, which represents a non-equatorial point of the parallelized 3-sphere \cite{What-666}.
Physically, it represents a rotation by angle ${2\theta_{{\widetilde{\bf a}}{\widetilde{\bf a}\,'}}}$ about
the ${\widetilde{\bf c}}$-axis. The limit ${{\widetilde{\bf a}\,'}\rightarrow{\widetilde{\bf a}}}$ then gives
${\left(+\,I\cdot{\widetilde{\bf a}}\,\right)\left(+\,I\cdot{\widetilde{\bf a}}\right)=-\,1}$,
which shows that unit bivectors square to ${-\,1}$. Moreover, in equation (\ref{14-2}) we have already seen the conjugate
limiting case; namely, ${(-\,I\cdot{\widetilde{\bf a}}\,)(+\,I\cdot{\widetilde{\bf a}}\,)=+\,1}$.
These two cases show that the pairs of numbers
${\{\,+\,1,\,-\,1\,\}}$ are intrinsic parts of the parallelized 3-sphere. They are in fact antipodal points of
one of its equators, representing the set of measurement results observed by our experimenters.

In statistical terms, however, the above variables are raw scores as opposed to standard scores \cite{scores-2}.
Recall that a standard score indicates how
many standard deviations the observation or datum is above or below the mean. If ${\rm x}$ is a raw (or unnormalized) score
and ${\overline{\rm x}}$ is its mean value, then the standard (or normalized) score, ${{\rm z}({\rm x})}$, is defined by
\begin{equation}
{\rm z}({\rm x})\,=\,\frac{{\rm x}\,-\,{\overline{\rm x}}}{\sigma({\rm x})}\,,
\end{equation}
where ${\sigma({\rm x})}$ is the standard deviation of ${\rm x}$. A standard score thus represents the distance between
a raw score and the population mean in the units of standard deviation, and
allows one to make comparisons of raw scores that come from very different sources. In other words, the
mean value of the standard score itself is always zero, with standard deviation unity.
In terms of these concepts the bivariate correlation between raw scores ${\rm x}$ and ${\rm y}$ is defined as
\begin{align}
{\cal E}({\rm x},\,{\rm y})\,&=\;\frac{\,{\displaystyle\lim_{\,n\,\gg\,1}}\left[{\displaystyle\frac{1}{n}}\,
{\displaystyle\sum_{i\,=\,1}^{n}}\,({\rm x}^i\,-\,{\overline{\rm x}}\,)\;
({\rm y}^i\,-\,{\overline{\rm y}}\,)\right]}{\sigma({\rm x})\;\sigma({\rm y})} \label{co} \\
&=\,\lim_{\,n\,\gg\,1}\left[\frac{1}{n}
\sum_{i\,=\,1}^{n}\,{\rm z}({\rm x}^i)\;{\rm z}({\rm y}^i)\right]. \label{stan}
\end{align}
It is vital to appreciate that covariance by itself---{\it i.e.}, the numerator of the equation (\ref{co}) by itself---does not
provide the correct measure of association between the raw scores, not the least because it depends on different units and scales
(or different scales of dispersion) that may have been used (advertently or inadvertently) in the measurements of such scores
\cite{scores-2}. Therefore, to arrive
at the correct measure of association between the raw scores one must either use equation (\ref{co}), with the product of standard
deviations in the denominator, or use covariance of the standardized variables, as in Eq.${\,}$(\ref{stan}).

These basic statistical concepts are crucial for understanding our local-realistic explanation of the EPR correlations.
As we saw above, the variables
${{\mathscr A}(\alpha,\,{\boldsymbol\mu})}$ and ${{\mathscr B}(\beta,\,{\boldsymbol\mu})}$ are pure binary numbers,
albeit occurring as antipodal points within a parallelized 3-sphere.
As random variables, however, they are products of two factors---one random and another non-random. For instance, within
${{\mathscr A}(\alpha,\,{\boldsymbol\mu})}$ the factor ${(\,+\,{\boldsymbol\mu}\cdot{\widetilde{\bf a}}\,)}$ is a random
bivector---a function of the hidden variable ${\boldsymbol\mu}$, whereas the factor
${(-\,I\cdot{\widetilde{\bf a}}\,)}$ is a non-random
bivector, independent of the hidden variable ${\boldsymbol\mu}$. Consequently, as a function, each of the numbers
${{\mathscr A}(\alpha,\,{\boldsymbol\mu})}$ and ${{\mathscr B}(\beta,\,{\boldsymbol\mu})}$ is generated with a
{\it different} standard deviation---{\it i.e.}, a {\it different} size of a typical error.
More specifically, the number ${{\mathscr A}(\alpha,\,{\boldsymbol\mu})}$ is generated with the standard deviation
${(-\,I\cdot{\widetilde{\bf a}}\,)}$, whereas the number ${{\mathscr B}(\beta,\,{\boldsymbol\mu})}$ is generated
with the standard deviation ${(+\,I\cdot{\widetilde{\bf b}}\,)}$. These deviations
can be calculated easily. Since errors in a linear relation such as (\ref{17-Joy}) propagate linearly,
the standard deviation of ${{\mathscr A}(\alpha,\,{\boldsymbol\mu})}$ is equal to ${(-\,I\cdot{\widetilde{\bf a}}\,)}$
times the standard deviation of ${(\,+\,{\boldsymbol\mu}\cdot{\widetilde{\bf a}}\,)}$ (which we write as ${\sigma({A})}$), and
similarly the standard deviation of ${{\mathscr B}(\beta,\,{\boldsymbol\mu})}$ is equal to ${(+\,I\cdot{\widetilde{\bf b}}\,)}$
times the standard deviation of ${(\,+\,{\boldsymbol\mu}\cdot{\widetilde{\bf b}}\,)}$ (which we write as ${\sigma({B})}$):
\begin{align}
\sigma({\mathscr A}\,)\,&=\,(\,-\,{I}\cdot{\widetilde{\bf a}}\,)\,\sigma({A}) \notag \\
\text{and}\;\;\sigma({\mathscr B}\,)\,&=\,(\,+\,{I}\cdot{\widetilde{\bf b}}\,)\,\sigma({B}).\label{asin23}
\end{align}
But since all bivectors we have been considering
are normalized to unity, and since the mean of ${(\,+\,{\boldsymbol\mu}\cdot{\widetilde{\bf a}}\,)}$ vanishes on the account
of ${\boldsymbol\mu}$ being a fair coin, its standard deviation is easy to calculate, and it turns out to be equal to unity:
\begin{align}
\sigma({A})\,&=\,\sqrt{\frac{1}{n}\sum_{i\,=\,1}^{n}\,\left|\left|\,A({\alpha},\,{\boldsymbol\mu}^i)\,-\,
{\overline{A({\alpha},\,{\boldsymbol\mu}^i)}}\;\right|\right|^2\,}\, \notag \\
&=\,\sqrt{\frac{1}{n}\sum_{i\,=\,1}^{n}\,
\left|\left|\,(\,+\,{\boldsymbol\mu}^i\cdot{\widetilde{\bf a}}\,)\,-\,0\,\right|\right|^2\,}\,=\,1,
\end{align}
where the last equality follows from the fact that all
bivectors ${(\,+\,{\boldsymbol\mu}^i\cdot{\widetilde{\bf a}}\,)}$ are normalized to unity.
Similarly, we find that the standard deviation ${\sigma({B})}$ is also equal to ${1}$. As a result, the standard deviation of
${{\mathscr A}(\alpha,\,{\boldsymbol\mu})}$ works out to be equal to ${(-\,I\cdot{\widetilde{\bf a}}\,)}$, and
the standard deviation of ${{\mathscr B}(\beta,\,{\boldsymbol\mu})}$ works out to be equal to ${(+\,I\cdot{\widetilde{\bf b}})}$.
Putting these two results together, we arrive at the following standard scores corresponding to the
raw scores (\ref{17-Joy}) and (\ref{18-Joy}):
\begin{align}
A(\alpha,\,{\boldsymbol\mu})&=\frac{\,{\mathscr A}(\alpha,\,{\boldsymbol\mu})\,-\,
{\overline{{\mathscr A}(\alpha,\,{\boldsymbol\mu})}}}{\sigma({\mathscr A})} \notag \\
\,&=\,\frac{\,{\mathscr A}(\alpha,\,{\boldsymbol\mu})\,-\,0\,}{(-\,I\cdot{\widetilde{\bf a}}\,)}
\,=\,(\,+\,{\boldsymbol\mu}\cdot{\widetilde{\bf a}}\,) \label{var-a}
\end{align}
and
\begin{align}
B(\beta,\,{\boldsymbol\mu})&=\frac{\,{\mathscr B}(\beta,\,{\boldsymbol\mu})\,-\,
{\overline{{\mathscr B}(\beta,\,{\boldsymbol\mu})}}}{\sigma({\mathscr B})} \notag \\
\,&=\,\frac{\,{\mathscr B}(\beta,\,{\boldsymbol\mu})\,-\,0\,}{(+\,I\cdot{\widetilde{\bf b}}\,)}
\,=\,(\,+\,{\boldsymbol\mu}\cdot{\widetilde{\bf b}}\,),\label{var-b}
\end{align}
where we have used the identities ${(+\,I\cdot{\widetilde{\bf a}})(-\,I\cdot{\widetilde{\bf a}})=+1}$
and ${(-\,I\cdot{\widetilde{\bf b}})(+\,I\cdot{\widetilde{\bf b}})=+1}$,
respectively (see Appendix for the formal basis of these results).

Not surprisingly, just like the raw scores ${{\mathscr A}(\alpha,\,{\boldsymbol\mu})}$ and
${{\mathscr B}(\beta,\,{\boldsymbol\mu})}$, these standard scores are also strictly
{\it local} variables: ${A(\alpha,\,{\boldsymbol\mu})}$ depends only on the freely chosen angle
${\alpha}$ and the initial state ${{\boldsymbol\mu}}$, and likewise ${B(\beta,\,{\boldsymbol\mu})}$ depends only on the
freely chosen angle ${\beta}$ and the initial state ${{\boldsymbol\mu}\,}$.
Moreover, despite appearances, ${\,A(\alpha,\,{\boldsymbol\mu})}$ and ${B(\beta,\,{\boldsymbol\mu})\,}$ are simply binary
measurement results, ${\pm\,1}$, albeit occurring within the
compact topology of the 3-sphere rather than the real line:
\begin{align}
S^3\supset S^2\ni A(\alpha,\,{\boldsymbol\mu})&\,=\,(\,+\,{\boldsymbol\mu}\cdot{\widetilde{\bf a}}\,)
\,=\, \pm\,1\;\,{\rm about}\;{\widetilde{\bf a}}\in{\rm I\!R}^3, \\
S^3\supset S^2\ni B(\beta,\,{\boldsymbol\mu})&\,=\,(\,+\,{\boldsymbol\mu}\cdot{\widetilde{\bf b}}\,)
\,=\, \pm\,1\;\,{\rm about}\;{\widetilde{\bf b}}\in{\rm I\!R}^3.
\end{align}
In fact, since the space of all bivectors
${(\,+\,{\boldsymbol\mu}\cdot{\widetilde{\bf a}}\,)}$ is isomorphic to the equatorial 2-sphere
contained within the 3-sphere \cite{Zulli}, each standard score ${A({\alpha},\,{\boldsymbol\mu})}$ of Alice---corresponding
to a predetermined element of reality of the photon---is uniquely identified with a definite point of this 2-sphere. In
other words, every predetermined
polarization of a given photon gives rise to a unique standard score ${A({\alpha},\,{\boldsymbol\mu})}$,
which in turn is unambiguously represented by a definite point of the equatorial 2-sphere, and likewise
for the standard scores ${B(\beta,\,{\boldsymbol\mu})}$ of Bob \cite{What-666}.

Now, since we have assumed that initially there was 50/50 chance between the right-handed and left-handed orientations of the
physical space---{\it i.e.}, equal chance between the initial states ${{\boldsymbol\mu}=+\,I}$ and ${{\boldsymbol\mu}=-\,I}$, the
expectation values of the local outcomes trivially work out to be
\begin{equation}
{\cal E}(\theta)\,=
\lim_{\,n\,\gg\,1}\left[\frac{1}{n}\sum_{i\,=\,1}^{n}\,
{\mathscr A}({\theta},\,{\boldsymbol\mu}^i)\,\right]\,=\,\overline{{\mathscr A}({\theta},\,{\boldsymbol\mu})}\,=\,0\,,
\end{equation}
where ${\theta=\alpha}$ or ${\beta}$. On the other hand, to determine the {\it correct} correlation
between the joint observations of Alice and Bob we must calculate covariance between the
standard scores ${{A}({\alpha},\,{\boldsymbol\mu})}$ and ${{B}({\beta},\,{\boldsymbol\mu})}$, not the raw scores:
\begin{equation}
{\cal E}({\alpha},\,{\beta})
=\lim_{\,n\,\gg\,1}\left[\frac{1}{n}\sum_{i\,=\,1}^{n}\, A({\alpha},\,{\boldsymbol\mu}^i)\;
B({\beta},\,{\boldsymbol\mu}^i)\right]. \label{jointAB}
\end{equation}
Conversely, the covariance of the raw scores will not give us the correct correlation between the observations of Alice and
Bob, because ${{\mathscr A}({\alpha},\,{\boldsymbol\mu})}$ and ${{\mathscr B}({\beta},\,{\boldsymbol\mu})}$ have been generated
with {\it different} scales of dispersion for each direction, and hence they are subject to {\it different} standard deviations
or random errors for each direction \cite{scores-2}.
On the other hand, the standard scores---by definition---have been standardized for all kinds
of possible random errors. Moreover, their product ${{A}({\alpha},\,{\boldsymbol\mu})\,{{B}({\beta},\,{\boldsymbol\mu})}}$---as
a non-equatorial point of the 3-sphere---represents a definite binary value:
\begin{equation}
S^3\ni A({\alpha},\,{\boldsymbol\mu})\,B({\beta},\,{\boldsymbol\mu}) \,=\, \pm\,1\;\,
\text{about a direction in}\;\,{\rm I\!R}^4.
\end{equation}
In fact, our 3-sphere is entirely made of such product points, each of definite value ${+\,1}$ or ${-\,1}$,
depending on its orientation defined by ${\boldsymbol\mu}$.
More precisely, the space of all such product points ${P=AB}$ is homotopic to a round 3-sphere:
${P_1^2+P_2^2+P_3^2+P_4^2=1}$ \cite{What-666}\cite{Zulli}. This
can be seen more clearly if we expand ${AB}$ using the identity (\ref{bi-identity}):
\begin{align}
A({\alpha},\,{\boldsymbol\mu})\,B({\beta},\,{\boldsymbol\mu})&=
(\,+\,{\boldsymbol\mu}\cdot{\widetilde{\bf a}}\,)
(\,+\,{\boldsymbol\mu}\cdot{\widetilde{\bf b}}\,) \notag \\
&=-\,{\widetilde{\bf a}}\cdot{\widetilde{\bf b}}\,-
\,{\boldsymbol\mu}\cdot(\,{\widetilde{\bf a}}\times{\widetilde{\bf b}}\,) \notag \\
&=-\cos2(\alpha-\beta)+({\boldsymbol\mu}\cdot{\bf e}_z) \, \sin2(\alpha-\beta). \label{circles}
\end{align}
Evidently, the product ${AB}$ describes a circle of points within ${S^3}$ \cite{Zulli}, each of definite value ${+\,1}$ or ${-\,1}$,
depending on ${\alpha}$, ${\beta}$, and ${\,{\boldsymbol\mu}\cdot{\bf e}_z\in S^2\subset S^3}$. Substituting this
parameterization of ${AB}$ into Eq.${\,}$(\ref{jointAB}), the correlation between the raw scores
${{\mathscr A}({\alpha},\,{\boldsymbol\mu})}$ and ${{\mathscr B}({\beta},\,{\boldsymbol\mu})}$
can now be easily seen to reproduce the corresponding quantum mechanical prediction:
\begin{align}
{\cal E}({\alpha},\,{\beta})
&=\,-\cos2(\alpha-\beta)\,+\lim_{\,n\,\gg\,1}\left[\frac{1}{n}\sum_{i\,=\,1}^{n}\,
 (\,{\boldsymbol\mu}^i\cdot{\bf e}_z) \, \sin2(\alpha-\beta)\right] \notag \\
&=\,-\cos2(\alpha-\beta)\,+\,0\,. \label{result}
\end{align}
Here the summation over microstates ${{\boldsymbol\mu}^i}$ has only formal significance since operationally ${{\bf e}_z}$
is a ``third'' direction, exclusive to both ${\bf a}$ and ${\bf b}$. If measurements along ${\bf a}$ and ${\bf b}$ have
yielded non-null polarizations, then simultaneous measurement along ${{\bf e}_z}$ could only yield a null result.

Suppose now we consider four possible polarization directions, ${\bf a}$, ${\bf a'}$, ${\bf b}$, and ${\bf b'}$.
Then, as is well known, the corresponding CHSH string of expectation values \cite{Clauser}, namely
\begin{equation}
{\cal E}({\bf a},\,{\bf b})\,+\,{\cal E}({\bf a},\,{\bf b'})\,+\,
{\cal E}({\bf a'},\,{\bf b})\,-\,{\cal E}({\bf a'},\,{\bf b'})\,,
\end{equation}
can be rewritten in terms of the products of our local functions
${{A}({\bf a},\,{\boldsymbol\mu})}$ and ${{B}({\bf b},\,{\boldsymbol\mu})}$ as
\begin{equation}
\lim_{\,n\,\gg\,1}\Bigg[\frac{1}{n}\sum_{i\,=\,1}^{n}\,\big\{
A_{\bf a}({\boldsymbol\mu}^i)\,B_{\bf b}({\boldsymbol\mu}^i)\,+\,
A_{\bf a}({\boldsymbol\mu}^i)\,B_{\bf b'}({\boldsymbol\mu}^i)\,+\,
A_{\bf a'}({\boldsymbol\mu}^i)\,B_{\bf b}({\boldsymbol\mu}^i)\,-\,
A_{\bf a'}({\boldsymbol\mu}^i)\,B_{\bf b'}({\boldsymbol\mu}^i)\big\}\Bigg]. \label{probnonint}
\end{equation}
But since the standard scores ${A_{\bf a}({\boldsymbol\mu})}$ and ${B_{\bf b}({\boldsymbol\mu})}$ represent
two independent points of ${S^3}$, we can take them to belong to two disconnected ``sections'' of ${S^3}$
({\it i.e.}, two disconnected 2-spheres within ${S^3}$), satisfying
\begin{equation}
\left[\,A_{\bf n}({\boldsymbol\mu}),\,B_{\bf n'}({\boldsymbol\mu})\,\right]\,=\,0\,
\;\;\;\forall\;\,{\bf n}\;\,{\rm and}\;\,{\bf n'}\,\in\,{\rm I\!R}^3,\label{com}
\end{equation}
which is operationally equivalent to assuming a null result along the direction ${{\bf n}\times{\bf n'}}$ exclusive
to both ${\bf n}$ and ${\bf n'}$.
If we now square the integrand of Eq.${\,}$(\ref{probnonint}), use the above commutation relations, and use the fact
that, by definition, all local functions square to unity (the algebra goes through even when the squares of the local
functions are allowed to be ${-1}$), then the absolute value of the CHSH string leads to the
following variance inequality:
\begin{align}
|{\cal E}({\bf a},\,{\bf b})\,+\,{\cal E}&({\bf a},\,{\bf b'})\,+\,
{\cal E}({\bf a'},\,{\bf b})\,-\,{\cal E}({\bf a'},\,{\bf b'})|\, \notag \\
&\leqslant\sqrt{\lim_{\,n\,\gg\,1}\left[\frac{1}{n}\sum_{i\,=\,1}^{n}\,\big\{\,4\,+\,\left[\,A_{\bf a}({\boldsymbol\mu}^i),\,
A_{\bf a'}({\boldsymbol\mu}^i)\,\right]\left[\,B_{\bf b'}({\boldsymbol\mu}^i),\,
B_{\bf b}({\boldsymbol\mu}^i)\,\right]\big\}\right]}.\label{yever}
\end{align}
Here the classical commutators ${\left[\,A_{\bf a}({\boldsymbol\mu}^i),\,A_{\bf a'}({\boldsymbol\mu}^i)\,\right]}$ and
${\left[\,B_{\bf b'}({\boldsymbol\mu}^i),\,B_{\bf b}({\boldsymbol\mu}^i)\,\right]}$ are simply geometric measures of
the parallelizing torsion within the 3-sphere \cite{What-666}.
Next, using the definitions (\ref{var-a}) and (\ref{var-b}) for the variables
${A_{\bf a}({\boldsymbol\mu})}$ and ${B_{\bf b}({\boldsymbol\mu})}$ and making a repeated use of
the identity ${(\,+\,{\boldsymbol\mu}\cdot{\widetilde{\bf a}}\,)(\,+\,{\boldsymbol\mu}\cdot{\widetilde{\bf a}\,'}\,)
\,=\,-\,{\widetilde{\bf a}}\cdot{\widetilde{\bf a}\,'}\,-\,{\boldsymbol\mu}
\cdot(\,{\widetilde{\bf a}}\times{\widetilde{\bf a}\,'}\,)}$, the above inequality can be simplified to
\begin{align}
|{\cal E}({\bf a},\,{\bf b})\,+\,{\cal E}&({\bf a},\,{\bf b'})\,+\,
{\cal E}({\bf a'},\,{\bf b})\,-\,{\cal E}({\bf a'},\,{\bf b'})|\, \notag \\
&\leqslant\,\sqrt{\lim_{\,n\,\gg\,1}\left[\frac{1}{n}\sum_{i\,=\,1}^{n}
\left\{4+4\big[\,-\,({\widetilde{\bf a}}\times{\widetilde{\bf a}'})\cdot({\widetilde{\bf b}'}
\times{\widetilde{\bf b}})\,-\,{\boldsymbol\mu}^i\cdot{\widetilde{\bf 0}}\;\big]\right\}\right]} \notag \\
&\leqslant\,2\,\sqrt{\,1-({\widetilde{\bf a}}\times{\widetilde{\bf a}\,'})
\cdot({\widetilde{\bf b}\,'}\times{\widetilde{\bf b}})\,}\,,\label{before-opppo-666}
\end{align}
where ${{\boldsymbol\mu}^i\cdot{\widetilde{\bf 0}}}$ is a null bivector. It is worth noting here that---because of the nullity of
${{\boldsymbol\mu}^i\cdot{\widetilde{\bf 0}}}$---the summation over the microstates ${{\boldsymbol\mu}^i}$
is not needed to arrive at the last inequality---the summation simply becomes redundant, bringing out the purely geometrical
character of the EPR correlations \cite{What-666} in addition to conforming the consistency of assumption (\ref{com}).
Moreover, using the definitions
(\ref{b-def}) for the vectors ${\widetilde{\bf a}}$ and ${\widetilde{\bf b}}$, the last inequality can be further simplified to
\begin{align}
|\,{\cal E}({\alpha},\,{\beta})\,+\,{\cal E}({\alpha},\,{\beta}\,')\,+\,
{\cal E}&({\alpha}\,',\,{\beta})\,-\,{\cal E}({\alpha}\,',\,{\beta}\,')\,|\, \notag \\
&\leqslant\,2\,\sqrt{\,1+\,\sin 2(\,\alpha\,-\,\alpha\,'\,)\,\sin 2(\,\beta\,-\,\beta\,'\,)}\,. \label{sinet}
\end{align}
It is now easy to see that a maximum violation of ${2\sqrt{2}}$ of the Bell-CHSH inequality can be achieved for the
angles ${({\alpha},\,{\alpha\,'},\,{\beta},\,{\beta}\,')=(0^{\circ},\,45^{\circ},\,22.5^{\circ},\,67.5^{\circ})}$,
precisely as observed in the experiments \cite{Aspect-666}\cite{Weihs-666}. On the other hand, by using
\begin{equation}
-1\leqslant\,\sin 2(\,\alpha\,-\,\alpha\,'\,)\,\sin 2(\,\beta\,-\,\beta\,'\,)\,\leqslant +1\,,
\end{equation}
the above inequality can be reduced to the form
\begin{equation}
\left|\,{\cal E}({\alpha},\,{\beta})\,+\,{\cal E}({\alpha},\,{\beta}\,')\,+\,
{\cal E}({\alpha}\,',\,{\beta})\,-\,{\cal E}({\alpha}\,',\,{\beta}\,')\,\right|\,\leqslant\,2\sqrt{2}\,,
\label{My-CHSH}
\end{equation}
which is precisely what is predicted by quantum mechanics, and observed in the experiments
\cite{Aspect-666}\cite{Weihs-666}.

We are now in a position to explain where exactly the illusion of quantum nonlocality stems from \cite{Bell-1964}.
The key to our explanation is the topology of the parallelized 3-sphere, or equivalently the algebra of rotations
in the physical space \cite{Clifford}\cite{What-666}.
As we saw in equation (\ref{circles}), topologically 3-sphere is simply a 2-sphere
worth of circles, but with a ``twist'' in the bundle \cite{Eguchi}\cite{Ryder}\cite{Can}.
That is to say, as an ${S^1}$ fiber bundle over ${S^2}$,
${S^3\not=S^2\times S^1}$. And this twist---analogous to the one in a m\"obius strip---is responsible for producing the right
combination of polarizations, namely ${++}$, ${--}$, ${+-}$, and ${-+}$, observed in the experiments
\cite{Aspect-666}\cite{Weihs-666}. To bring this out explicitly, we
have modeled the physical space, not as ${{\rm I\!R}^3}$, but ${S^3}$, in terms of coordinate-free geometry of points
and planes introduced by Grassmann \cite{Clifford}. This, in turn, has allowed us to introduce the raw scores observed
by Alice and Bob as ${{\mathscr A}(\alpha,\,{\boldsymbol\mu})=(-\,I\cdot{\widetilde{\bf a}}\,)
\,(\,+\,{\boldsymbol\mu}\cdot{\widetilde{\bf a}}\,)=\pm\,1}$ and ${{\mathscr B}(\beta,\,{\boldsymbol\mu})
=(+\,I\cdot{\widetilde{\bf b}}\,)\,(\,+\,{\boldsymbol\mu}\cdot{\widetilde{\bf b}}\,)=\pm\,1}$, respectively,
with the random factors ${(\,+\,{\boldsymbol\mu}\cdot{\widetilde{\bf a}}\,)}$ and
${(\,+\,{\boldsymbol\mu}\cdot{\widetilde{\bf b}}\,)}$ representing the photon polarizations,
and the fixed factors ${(-\,I\cdot{\widetilde{\bf a}}\,)}$ and ${(+\,I\cdot{\widetilde{\bf b}}\,)}$ representing the
analyzers used by Alice and Bob to measure these polarizations. Furthermore, we have taken the randomness
${{\boldsymbol\mu}=+\,I}$ or ${-\,I}$ shared by Alice and Bob
to be the initial orientation (or handedness) of the entire physical space,
or equivalently that of a 3-sphere. Consequently, once ${\boldsymbol\mu}$ is given as an initial state,
the polarizations
along all directions chosen by Alice and Bob would have the same value, because ${\boldsymbol\mu}$ completely fixes the sense
of bivectors ${{\boldsymbol\mu}\cdot{\bf n}}$ belonging to ${S^2\subset S^3}$, regardless of direction. 
However, and this is an important point, the polarization  ${(\,+\,{\boldsymbol\mu}\cdot{\widetilde{\bf a}}\,)}$ observed
by Alice is measured with respect to the analyzer ${(-\,I\cdot{\widetilde{\bf a}}\,)}$, whereas the polarization 
${(\,+\,{\boldsymbol\mu}\cdot{\widetilde{\bf b}}\,)}$ observed by Bob is measured with respect to the analyzer
${(+\,I\cdot{\widetilde{\bf b}}\,)}$.

To appreciate the significance of the last point, let us begin with the scenario in which Alice and Bob are oblivious
of each other's existence. Then, as is evident from equations (\ref{17-Joy}) and (\ref{18-Joy}), their measurement results
would depend only on the sense of ${\boldsymbol\mu}$. Consequently, Alice, for example, would be justified in concluding from
${{\mathscr A}(\alpha,\,{\boldsymbol\mu})=+\,1}$ that the local
handedness associated with the points of ${S^3}$ is of counterclockwise
variety, since ${(-\,I\cdot{\widetilde{\bf a}}\,)\,(\,+\,{\boldsymbol\mu}\cdot{\widetilde{\bf a}}\,)=+\,1}$ holds for any direction
${{\widetilde{\bf a}}}$ chosen by Alice if ${{\boldsymbol\mu}=+\,I}$; and, ${\boldsymbol\mu}$, as we noted, defines a
consistent sense of handedness over the whole of ${S^3}$ for each run of the experiment.
Similarly, Bob too would be justified in concluding from
${{\mathscr B}(\beta,\,{\boldsymbol\mu})=-\,1}$ that the points of the 3-sphere are oriented in the counterclockwise sense,
since ${(+\,I\cdot{\widetilde{\bf b}}\,)\,(\,+\,{\boldsymbol\mu}\cdot{\widetilde{\bf b}}\,)=-\,1}$ holds for any direction
${{\widetilde{\bf b}}}$ chosen by Bob if ${{\boldsymbol\mu}=+\,I}$. Furthermore, these conclusions are unproblematic in the
important case of ``perfect anti-correlation''---i.e., when ${\widetilde{\bf b}}$ happens to be equal to ${\widetilde{\bf a}}$.
But since we have ${{\mathscr A}=+\,1}$ and ${{\mathscr B}=-\,1}$ in this case, this is, in fact, the case ``${+-}$''
observed in the experiments. Analogously, for the initial orientation
${{\boldsymbol\mu}=-\,I}$ the results of Alice and Bob would be ${{\mathscr A}=-\,1}$ and ${{\mathscr B}=+\,1}$, and that would
be the case ``${-+}$'' observed in the experiments (with equal probability). Note, however, that so far we have no indication of
``nonlocality.''

But now suppose we let ${{\widetilde{\bf b}}\,\rightarrow\,-\,{\widetilde{\bf a}}}$. Then
${(+\,I\cdot{\widetilde{\bf b}}\,)}$ becomes ${\{+\,I\cdot(-\,{\widetilde{\bf a}})\}=(-\,I\cdot{\widetilde{\bf a}}\,)}$,
and---unlike in the previous case---the analyzers used by
Alice and Bob become mathematically identical, with the corresponding raw scores
${\mathscr A}$ and ${\mathscr B}$ generated with the same scale of dispersion for each direction. As a result, now it
would be impossible for Alice and Bob to observe opposite polarizations without violating the consistency
of handedness defined by ${\boldsymbol\mu}$ over the whole of ${S^3}$. Thus, if Alice's result turns out to be
${(-\,I\cdot{\widetilde{\bf a}}\,)\,(\,+\,{\boldsymbol\mu}\cdot{\widetilde{\bf a}}\,)}$ as before, then Bob's result must also
be ${(-\,I\cdot{\widetilde{\bf a}}\,)\,(\,+\,{\boldsymbol\mu}\cdot{\widetilde{\bf a}}\,)}$, albeit for the
detection of polarization along the angle ${\alpha+\frac{\pi}{2}}$. In other words, the
consistency of handedness throughout the 3-sphere  necessitates that the only possible results Alice and Bob could
observe in this case
are either ${{\mathscr A}=+\,1}$ and ${{\mathscr B}=+\,1}$, or ${{\mathscr A}=-\,1}$ and ${{\mathscr B}=-\,1}$,
depending on whether the initial orientation for the run has been ${{\boldsymbol\mu}=+\,I}$ or ${-\,I}$. These are, then,
the cases ``${++}$'' and ``${--}$'' observed in the experiments (again, with equal probabilities). 

Following Bell \cite{Bell-1964}, such instantaneous changes in the relative measurement outcomes observed at remote stations
are usually taken as evidence of ``quantum nonlocality'' \cite{Nature-666}. In our spherical model of the physical space,
however, such changes are determined
entirely by the intrinsic geometrical and topological structures of the 3-sphere (or equivalently by the algebra of rotations
in the physical space) without requiring any form of communication between Alice and Bob.
To appreciate this in detail,
let us express the results of Alice and Bob as limiting cases of two quaternions constituting the 3-sphere:
\begin{equation}
S^3\ni {\mathscr A}(\alpha,\,{\boldsymbol\mu})\,=\,\lim_{{\widetilde{\bf a}}\,'\rightarrow{\widetilde{\bf a}}}
\big[(-\,I\cdot{\widetilde{\bf a}}\,)\,(\,+\,{\boldsymbol\mu}\cdot{\widetilde{\bf a}}\,'\,)\big]\,=\,+\,\lambda
\end{equation}
and
\begin{equation}
S^3\ni {\mathscr B}(\beta,\,{\boldsymbol\mu})\,=\,\lim_{{\widetilde{\bf b}}\,'\rightarrow{\widetilde{\bf b}}}
\big[(+\,I\cdot{\widetilde{\bf b}}\,)\,(\,+\,{\boldsymbol\mu}\cdot{\widetilde{\bf b}}\,'\,)\big]\,=\,-\,\lambda\,,
\end{equation}
where ${\lambda=\pm\,1}$, with ${{\boldsymbol\mu}=\lambda\,I}$, and we assume that the angles
${\theta_{{\widetilde{\bf a}}{{\widetilde{\bf a}}\,'}}}$ and ${\theta_{{\widetilde{\bf b}}{{\widetilde{\bf b}}\,'}}}$
between ${\widetilde{\bf a}}$ and ${{\widetilde{\bf a}}\,'}$ and ${\widetilde{\bf b}}$ and ${{\widetilde{\bf b}}\,'}$,
respectively, are infinitesimally small. Moreover, as we saw in Eq.${\,}$(\ref{quat}),
the quaternion ${(+\,I\cdot{\widetilde{\bf a}}\,)\,(\,+\,{\boldsymbol\mu}\cdot{\widetilde{\bf a}}\,'\,)}$ can be decomposed as
\begin{align}
(+\,I\cdot{\widetilde{\bf a}}\,)\,(\,+\,{\boldsymbol\mu}\cdot{\widetilde{\bf a}}\,'\,)
&=\,-\,\lambda\,
\{\,{\widetilde{\bf a}}\cdot{\widetilde{\bf a}\,'}\,+\,I\cdot(\,{\widetilde{\bf a}}\times{\widetilde{\bf a}\,'\,})\,\}
\notag \\
&=\,-\,\lambda\,\{\,\cos\theta_{{\widetilde{\bf a}}{\widetilde{\bf a}\,'}}\,+\,\left(\,I\cdot{\widetilde{\bf c}}
\,\right)\,\sin\theta_{{\widetilde{\bf a}}{\widetilde{\bf a}\,'}}\,\}\,, \label{quat-expan}
\end{align}
with ${{\widetilde{\bf c}}={\widetilde{\bf a}}\times{\widetilde{\bf a}\,'}/|{\widetilde{\bf a}}\times{\widetilde{\bf a}\,'}|}$.
Now suppose the pair ${{\widetilde{\bf b}}\,{{\widetilde{\bf b}}\,'}}$ of vectors starts out being aligned with the pair
${{\widetilde{\bf a}}\,{{\widetilde{\bf a}}\,'}}$,
and then gets rotated counterclockwise by angle ${\theta_{{\widetilde{\bf a}}{\widetilde{\bf b}}}}$
relative to the pair ${{\widetilde{\bf a}}\,{{\widetilde{\bf a}}\,'}}$, about the ${\widetilde{\bf c}}$-axis.
Such a rotation within ${S^3}$ can be described {\it intrinsically}, as
\begin{align}
(+\,I\cdot{\widetilde{\bf b}}\,)\,(\,+\,{\boldsymbol\mu}\cdot{\widetilde{\bf b}}\,'\,)\,&=\,
{\cal R}_{{\widetilde{\bf a}}{\widetilde{\bf b}}}\,
\left\{(\,+\,I\cdot{\widetilde{\bf a}}\,)\,(\,+\,{\boldsymbol\mu}\cdot{\widetilde{\bf a}}\,'\,)\right\} \notag \\
&=\,-\,\lambda\,\{\,\cos\left(\theta_{{\widetilde{\bf a}}{\widetilde{\bf b}}}\,+\,
\theta_{{\widetilde{\bf a}}{\widetilde{\bf a}\,'}}\right)
\,+\,\left(\,I\cdot{\widetilde{\bf c}}\,\right)\,\sin\left(\theta_{{\widetilde{\bf a}}{\widetilde{\bf b}}}\,+\,
\theta_{{\widetilde{\bf a}}{\widetilde{\bf a}\,'}}\right)\}\,,
\end{align}
where ${{\cal R}_{{\widetilde{\bf a}}{\widetilde{\bf b}}}\,=\,\exp\left\{(\,I\cdot{\widetilde{\bf c}}\,)\,
\theta_{{\widetilde{\bf a}}{\widetilde{\bf b}}}\right\}\,=\,{{\widetilde{\bf a}}\,{\widetilde{\bf b}}}\,}$
is the rotor that parallel transports the quaternion
${(+\,I\cdot{\widetilde{\bf b}}\,)\,(\,+\,{\boldsymbol\mu}\cdot{\widetilde{\bf b}}\,'\,)}$
from ``${\,}$point ${\widetilde{\bf a}\,}$'' to ``${\,}$point ${\widetilde{\bf b}\,}$'' within the
3-sphere \cite{Clifford}\cite{What-666}\cite{Eguchi}.
But using the identity
\begin{equation*}
\cos\left(\theta_{{\widetilde{\bf a}}{\widetilde{\bf b}}}\,+\,
\theta_{{\widetilde{\bf a}}{\widetilde{\bf a}\,'}}\right)
=\cos\theta_{{\widetilde{\bf a}}{\widetilde{\bf b}}}\,\cos\theta_{{\widetilde{\bf a}}{\widetilde{\bf a}\,'}}\,-\,
\sin\theta_{{\widetilde{\bf a}}{\widetilde{\bf b}}}\,\sin\theta_{{\widetilde{\bf a}}{\widetilde{\bf a}\,'}}\,,
\end{equation*}
and noting that ${I\cdot{\widetilde{\bf c}}}$ in the limit ${{\widetilde{\bf a}\,'}\rightarrow{\widetilde{\bf a}}}$
reduces to a null bivector, it is easy to deduce from this rotation that
\begin{align}
{\mathscr B}(\beta,\,{\boldsymbol\mu})\,&=\,\lim_{{\widetilde{\bf b}\,'}\rightarrow{\widetilde{\bf b}}}
\big[\,(+\,I\cdot{\widetilde{\bf b}}\,)\,(\,+\,{\boldsymbol\mu}\cdot{\widetilde{\bf b}}\,'\,)\,\big]\,=\,-\,\lambda \notag \\
\,&\longrightarrow\,
\lim_{{\widetilde{\bf a}\,'}\rightarrow{\widetilde{\bf a}}}\big[\,{\cal R}_{{\widetilde{\bf a}}{\widetilde{\bf b}}}
\left\{(+\,I\cdot{\widetilde{\bf a}}\,)\,(\,+\,{\boldsymbol\mu}\cdot{\widetilde{\bf a}}\,'\,)\right\}\big]
\,=\,-\,\lambda\,\cos\theta_{{\widetilde{\bf a}}{\widetilde{\bf b}}}\,=\,-\,\lambda\,\cos2(\alpha-\beta)\,.\label{un}
\end{align}
This is a probabilistic prediction that says that the relative frequency of occurring the result
${{\mathscr B}(\beta,\,{\boldsymbol\mu})=-\,\lambda}$ will vary with the cosine of the
angle ${\theta_{{\widetilde{\bf a}}{\widetilde{\bf b}}}}$. In particular,
as ${{\widetilde{\bf b}}={\widetilde{\bf a}}\,\rightarrow\,-\,{\widetilde{\bf a}}}$,
the result ${{\mathscr B}(\beta,\,{\boldsymbol\mu})}$ will change from ${-\,\lambda}$ to ${+\,\lambda}$,
and consequently the value of the product ${{\mathscr A}(\alpha,\,{\boldsymbol\mu})\,{\mathscr B}(\beta,\,{\boldsymbol\mu})}$
will change from ${-1}$ to ${+1}$, just as we discussed above. This, then, is a manifestation of the ``twist'' in the Hopf
fibration of the 3-sphere mentioned above \cite{Eguchi}\cite{Ryder}\cite{Can}. By contrast, had we modeled the physical
space as ${{\rm I\!R}^3}$ instead of ${S^3}$, the same result would have appeared as a nonlocal effect.

In the previous paragraphs we investigated the cases
${{\widetilde{\bf b}}\,\rightarrow\,{\widetilde{\bf a}}}$ and ${{\widetilde{\bf b}}\,\rightarrow\,-\,{\widetilde{\bf a}}}$
in some detail. For the general case
${{\widetilde{\bf b}}\not={\widetilde{\bf a}}}$ it is convenient to view the analyzers
${(-\,I\cdot{\widetilde{\bf a}}\,)}$ and ${(+\,I\cdot{\widetilde{\bf b}}\,)}$
as two different scales of measurements, as we have done above.
Accordingly, we may think of polarization ${(\,+\,{\boldsymbol\mu}\cdot{\widetilde{\bf a}}\,)}$ observed by Alice as scaled
by ${(-\,I\cdot{\widetilde{\bf a}}\,)}$ to obtain the raw score ${{\mathscr A}(\alpha,\,{\boldsymbol\mu})=\pm\,1}$,
and likewise polarization ${(\,+\,{\boldsymbol\mu}\cdot{\widetilde{\bf b}}\,)}$ observed by Bob as
scaled by ${(+\,I\cdot{\widetilde{\bf b}}\,)}$ to obtain the raw score ${{\mathscr B}(\beta,\,{\boldsymbol\mu})=\pm\,1}$.
But since ${(-\,I\cdot{\widetilde{\bf a}}\,)}$ and ${(+\,I\cdot{\widetilde{\bf b}}\,)}$ are two {\it different} scales measuring
the same orientation of ${S^3}$, the correct correlation between the raw scores can only be determined by comparing
the corresponding standard scores, as we have done in Eqs.${\,}$(\ref{jointAB}) to (\ref{My-CHSH}). Alternatively
(but equivalently), the correct correlation between the raw scores
can be determined by calculating their covariance divided by
the product of their standard deviations, as explained in Eq.${\,}$(\ref{co}), with the product of the standard deviations
in our case being ${(-\,I\cdot{\widetilde{\bf a}}\,)(+\,I\cdot{\widetilde{\bf b}}\,)={\widetilde{\bf a}}\,{\widetilde{\bf b}}}$
(see also \cite{disproof}). This product however is precisely the rotor that quantifies the twist in the
fibration of ${S^3}$, as in Eq.${\,}$(\ref{un}) above (see also Ref.${\,}$\cite{Eguchi}).
Its value varies from ${+1}$ for ${{\widetilde{\bf b}}={\widetilde{\bf a}}}$ to ${-1}$ for
${{\widetilde{\bf b}}=-\,{\widetilde{\bf a}}}$ and back, producing the correct combination of
probabilities for detection of polarizations
${++}$, ${--}$, ${+-}$, and ${-+}$ observed in the experiments.
Consequently, in accordance with our results (\ref{result}) and (\ref{un}),
when the raw scores ${{\mathscr A}=\pm\,1}$ and ${{\mathscr B}=\pm\,1}$ are compared in practice by coincidence counts
\cite{Aspect-666}\cite{Weihs-666}, the normalized expectation value of their product will inevitably yield
\begin{align}
{\cal E}({\alpha},\,{\beta})\,&=\,\frac{\Big[C_{++}({\alpha},\,{\beta})\,+\,C_{--}({\alpha},\,{\beta})
\,-\,C_{+-}({\alpha},\,{\beta})\,-\,C_{-+}({\alpha},\,{\beta})\Big]}{\Big[C_{++}({\alpha},\,{\beta})\,+\,C_{--}({\alpha},\,{\beta})
\,+\,C_{+-}({\alpha},\,{\beta})\,+\,C_{-+}({\alpha},\,{\beta})\Big]}\, \notag \\
&=\lim_{\,n\,\gg\,1}\left[\frac{1}{n}\sum_{i\,=\,1}^{n}\, {\mathscr A}({\alpha},\,{\boldsymbol\mu}^i)\;
{\mathscr B}({\beta},\,{\boldsymbol\mu}^i)\right] \notag \\
&=\,-\cos2(\alpha-\beta)\,, \label{calcul}
\end{align}
where ${C_{+-}({\alpha},\,{\beta})}$ etc. represent the number of joint occurrences of detections
${+\,1}$ along ${\bf a}$ and ${-\,1}$ along ${\bf b}$ etc.

Now the important feature of the experiments performed in Orsay \cite{Aspect-666} and Innsbruck \cite{Weihs-666}
was that the measurement settings at the remote polarizers were
changed {\it during} the flight of the two particles, thus removing any possibility of communication between the two ends.
The Innsbruck team precluded the possibility of communication between the polarizers by using random and ultrafast switching
of the orientations of the polarizers. On each side of the experiment a local computer registered the polarizer orientation
and the result of each measurement, with timing monitored by an atomic clock, and the data was gathered and compared for
correlation only after the end of each run. These achievements prompted the principal
investigator of the Orsay experiment to make the following comment:
\begin{quote}
I suggest we take the point of view of an external observer, who collects the data from the two
distant stations at the end of the experiment, and compares the two series of results.
This is what the Innsbruck team has done. Looking at the data a posteriori, they found that the
correlation immediately changed as soon as one of the polarizers was switched, without any delay
allowing for signal propagation${\,}$... \cite{Nature-666}.
\end{quote}
But given the explanation above this immediate change in correlation should be no more puzzling than the sudden change in
the color of Dr. Bertlmann's sock discussed by Bell \cite{Sock}. In fact, in our opinion classical men like Hamilton,
Grassmann, Clifford, and Hopf would not have been puzzled by this change at all. They would have explained the
results observed in the experiments in exactly the same local terms as we have explained them here; namely, in terms of
the algebra of rotations in the physical space ${S^3}$, otherwise known as Clifford algebra
\cite{Clifford}\cite{What-666}\cite{Can}\cite{disproof}.

\acknowledgments

I wish to thank Abner Shimony for suggesting that I should explain the results of Orsay and Innsbruck experiments.
I also wish to thank the Foundational Questions Institute (FQXi) for supporting this work through a Mini-Grant, and
Heine Rasmussen for inspiring the appendix.

\begin{appendices}

\renewcommand*\thesection{\Alph{section}}

\section{${\!\!\!\!\!\!\!\!\!}$ppendix${:\;\,\;}$Error Propagation Within a Parallelized 3-Sphere}

{\parindent 0pt

In this appendix we spell out the statistical basis of the results (\ref{var-a})
and (\ref{var-b}) in more detail. To this end (recalling that a unit 3-sphere is
homeomorphic to a set of unit quaternions), consider a probability density function
${P({\bf q}):S^3\rightarrow\,[{\hspace{1pt}}0,\,1]}$ of random quaternions over ${S^3}$:}
\begin{equation}
P({\bf q})\,=\,\frac{1}{\sqrt{2\pi\left|\left|{\hspace{1pt}}\sigma({\bf q})\right|\right|^2\,}\;}\,
\exp\left\{-\,\frac{\,\left|\left|\,{\bf q}-m({\bf q})\right|\right|^2}{2\,
\left|\left|{\hspace{1pt}}\sigma({\bf q})\right|\right|^2}\right\}.\label{probdensi-8}
\end{equation}
It is a matter of indifference whether the distribution of ${\bf q}$ so chosen happens to be normal or not \cite{scores-2}.
Here ${{\bf q}={\mathscr A}+{\bf w}=\text{scalar}+\text{bivector}}$ is a generic quaternion within ${S^3}$
whose mean value ${m({\bf q})}$ is defined as
\begin{equation}
m({\bf q})\,=\,\frac{1}{n}\sum_{i\,=\,1}^{n}\,{\bf q}^i\,
\end{equation}
and whose standard deviation ${\sigma({\bf q})}$ (with ${||{\bf q}||^2={\bf q}^\dagger{\bf q}}$) is defined as
\begin{equation}
\sigma({\bf q})\,=\,\sqrt{\frac{1}{n}\sum_{i\,=\,1}^{n}\,\left|\left|\,{\bf q}^i\,-\,m({\bf q})\right|\right|^2\,}\,.\label{esbo-z}
\end{equation}
Needless to say, these definitions remain valid for the limiting cases of the random variable ${{\bf q}\in S^3}$
when ${{\bf q}={\mathscr A}\in[-1,\,+1]}$ and ${{\bf q}={\bf w}\in S^2}$.
Now let ${{\bf w}=p\,{\boldsymbol\mu}\cdot{\bf n}\in S^3}$ be a random bivector with
${||{\boldsymbol\mu}\cdot{\bf n}||^2=1}$ and ${p\in[{\hspace{1pt}}0,\,1]}$,
where ${{\boldsymbol\mu}=\lambda\,I}$ is the indefinite volume form as before and
${{\bf n}\in{\rm I\!R}^3}$ is a unit vector. Then, using equation (\ref{esbo-z}),
it is easy to verify that, in general, the mean ${m({\bf w})}$ of ${\bf w}$ and the standard deviation
${\sigma({\bf w})}$ of ${\bf w}$, respectively, would be a bivector and a scalar:
\begin{align}
m({\bf w}) & \,=\;\text{a bivector}\;\;\;\;\;\;\;\; \notag \\
\text{and}\;\;\;\sigma({\bf w})& \,=\;\text{a scalar}.
\end{align}

Suppose now we consider another non-random bivector, ${{\bf v}\!=\!I\cdot{\bf n}}$, and
define a scalar number ${\,-1 \leq {\mathscr A} = {\bf v}\,{\bf w} \leq +1\,}$, so that
${m({\mathscr A})\geq 0}$. Since ${\bf v}$ is a non-random
bivector, errors generated within ${\mathscr A}$ by the random process ${p\lambda}$ stem
entirely from the random bivector ${\bf w}$, and propagate linearly. In other words,
the standard deviations within the random number ${\mathscr A}$ due to the random
process ${p\lambda}$ is given by
\begin{equation}
\sigma({\mathscr A})={\bf v}\,\sigma({\bf w}).
\end{equation}
But since ${\sigma({\bf w})}$ is a scalar, the typical error 
${\sigma({\mathscr A})}$ generated within ${\mathscr A}$ due to the
random process ${p\lambda}$ is a bivector \cite{disproof}. The standardized variable
(which must be used to compare the raw scores ${\mathscr A}$ with other raw
scores ${{\mathscr B}\,}$) is thus also a bivector: ${A:={\mathscr A}/{\sigma({\mathscr A})}=\text{scalar}\times{\bf w}}$.

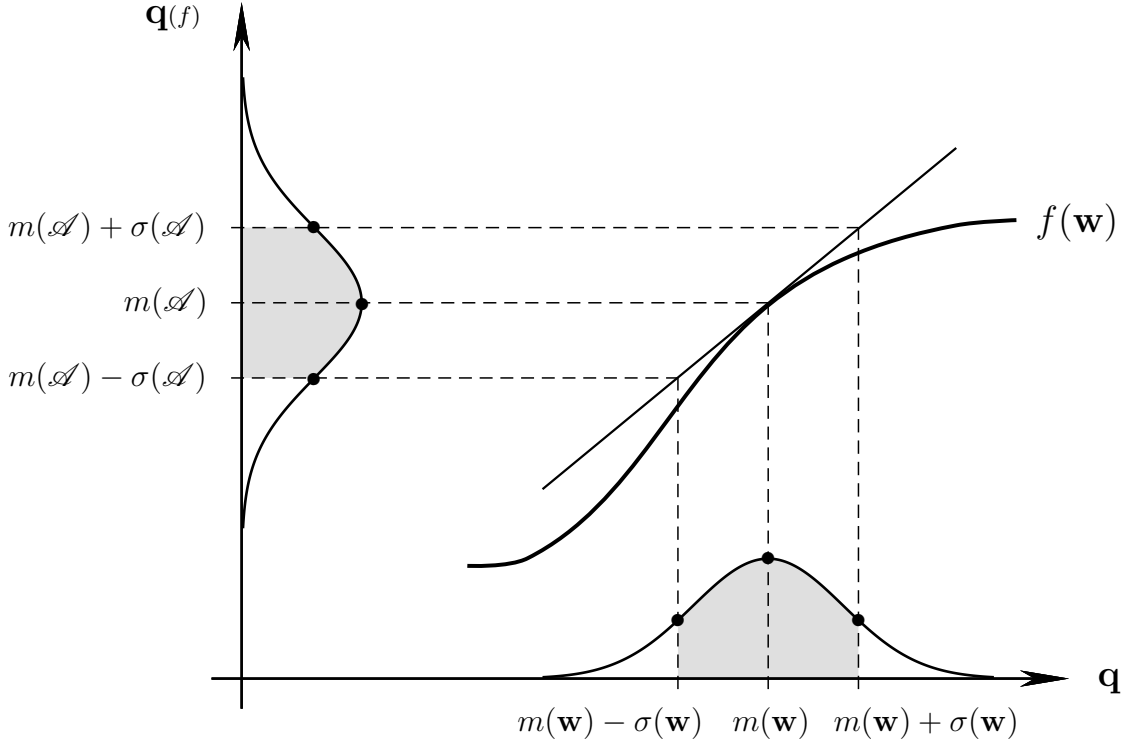
\begin{figure}
\hrule
%\vspace{-0.2cm}
\scalebox{1}{
\psset{yunit=2cm,xunit=2}
\begin{pspicture}(-0.67,-0.5)(5,5)

\uput[0](-0.73,4.4){{\Large ${\bf q}$}${(f)}$}

%\uput[0](-0.55,4.4){\Large ${\approx {\mathscr A}+{\bf 0}}$}

%\uput[0](-1.05,4.4){\Large ${{\mathscr A}+{\hspace{1pt}}{\bf 0}}$}

\uput[0](-1.64,3.0){\large ${m({\mathscr A})+\sigma({\mathscr A})}$}

\uput[0](-0.87,2.5){\large ${m({\mathscr A})}$}

\uput[0](-1.64,2.0){\large ${m({\mathscr A})-\sigma({\mathscr A})}$}

\uput[-90](5.765,0.15){\Large ${\bf q}$}

%\uput[-90](5.765,0.15){\Large ${\approx 0+{\bf w}}$}

%\uput[-90](6.05,0.19){\Large ${{\bf w}+0}$}

\uput[-90](5.55,3.25){\Large ${f}$({\bf w})}

\uput[-90](2.45,-0.1){\large ${m({\bf w})-\sigma({\bf w})}$}

\uput[-90](3.52,-0.1){\large ${m({\bf w})}$}

\uput[-90](4.55,-0.1){\large ${m({\bf w})+\sigma({\bf w})}$}

%\uput[-90](4.25,4.6){\LARGE ${S^{\,3}}$}

%\uput[-90](3.25,4.6){\LARGE ${S^{\,3}\times S^{\,3}}$}

\pscustom[linestyle=none,fillstyle=solid,fillcolor=gray!25]{%
\psGauss[sigma=0.5,mue=3.5,linewidth=1pt]{2.9}{4.1}
      \psline(4.1,0.0)(2.9,0.0)}

\psGauss[sigma=0.5,mue=3.5,linewidth=1pt]{2.0}{5.0}

\uput[-90](2.9,0.519){\large ${\bullet}$}

\uput[-90](3.5,0.931){\large ${\bullet}$}

\uput[-90](4.1,0.519){\large ${\bullet}$}

\psline[linewidth=0.2mm,linestyle=dashed]{-}(2.9,-0.07)(2.9,2.0)

\psline[linewidth=0.2mm,linestyle=dashed]{-}(4.1,-0.07)(4.1,3.0)

\begin{rotate}{-90}
\pscustom[linestyle=none,fillstyle=solid,fillcolor=gray!25]{%
\psGauss[sigma=0.5,mue=-2.5,linewidth=1pt,fillstyle=solid,fillcolor=gray!25]{-3.00}{-2.0}%
      \psline(-2.0,0.0)(-3.0,0.0)}
\end{rotate}

\psaxes[labels=none,ticksize=0pt,arrowinset=0.3,arrowsize=3pt 4,arrowlength=3]{->}(0,0)(-0.2,-0.2)(5.5,4.5)

\psline[linewidth=0.2mm,linestyle=dashed]{-}(-0.07,2.0)(2.9,2.0)

\psline[linewidth=0.2mm,linestyle=dashed]{-}(-0.07,3.0)(4.1,3.0)

\begin{rotate}{-90}
\psGauss[sigma=0.5,mue=-2.5,linewidth=1pt]{-4.0}{-1.0}%
\end{rotate}

\uput[-90](0.48,3.134){\large ${\bullet}$}

\uput[-90](0.8,2.625){\large ${\bullet}$}

\uput[-90](0.48,2.125){\large ${\bullet}$}

\psline[linewidth=0.3mm]{-}(2.0,1.26)(4.75,3.53)

\psline[linewidth=0.2mm,linestyle=dashed]{-}(3.5,-0.07)(3.5,2.5)

\psline[linewidth=0.2mm,linestyle=dashed]{-}(-0.07,2.5)(3.5,2.5)

\pscurve[linewidth=1.5pt]{-}(1.5,0.75)(1.9,0.8)(3.5,2.483)(4.0,2.79)(4.75,3.015)(5.15,3.05)

\end{pspicture}}
\vspace{0.25cm}
\hrule
\caption{Propagation of error within a parallelized 3-sphere.\break}
\label{fig-88}
\vspace{0.3cm}
\hrule
\end{figure}

As straightforward as it is, the above conclusion may seem rather unusual.
It is important to recall, however, that in geometric algebra scalars and bivectors are treated
on equal footing. They both behave as real-valued c-numbers, albeit of different grades \cite{What-666}.
To appreciate the consistency and naturalness of the above conclusion, let
\begin{equation}
{\mathscr A}\,=\,f({\bf w})\,=\,{\bf v}\,{\bf w}\label{defofa-0}
\end{equation}
be a continuous random scalar generated by the geometric product of the two bivectors ${\bf v}$ and ${\bf w}$, as before.
The natural question then is: How does a typical error in ${\bf w}$ governed by the probability density
(\ref{probdensi-8})---which can be represented by the 68\% probability interval
\begin{equation}
\left[\,m({\bf w})-\sigma({\bf w}),\;m({\bf w})+\sigma({\bf w})\,\right]\label{int-1no}
\end{equation}
as shown in the Fig.${\,}$(\ref{fig-88})---propagate from the random bivector ${\bf w}$ to the random
scalar ${\mathscr A}$, through the function ${f({\bf w}) = {\bf v}\,{\bf w}}$? To answer this question
we note that the two end points of the interval (\ref{int-1no}) represent two non-equatorial points,
${{\bf q}^-}$ and ${{\bf q}^+}$, of the parallelized 3-sphere, which is a Riemannian manifold. The
geometro-algebraic distance between the points ${{\bf q}^-}$ and ${{\bf q}^+}$ can therefore be defined as
\begin{equation}
d{\hspace{-1pt}}\left({\bf q}^-\!,\,{\bf q}^+\right)
\,=\,\left({\bf q}^- -\,{\bf q}^+\right)\times \text{sign}\!\left({\bf q}^- -\,{\bf q}^+\right).
\end{equation}
Moreover, from the definition (\ref{defofa-0}) of ${\mathscr A}$ and a first-order Taylor expansion of
the function ${f({\bf w})}$ about the point ${{\bf w} = m({\bf w})}$ we obtain
\begin{equation}
{\mathscr A}\,=\,f(m({\bf w}))\,+\,\frac{\partial f}{\partial {\bf w}}\bigg|_{{\bf w}\,=\;m({\bf w})}
({\bf w}\,-\,m({\bf w}))\,+\,\dots \label{Tylore}
\end{equation}
Now it is evident that the slope ${{\partial f}/{\partial {\bf w}}={\bf v}}$ of this line is a constant.
Therefore the mean ${m({\mathscr A})}$ and the standard deviation ${\sigma({\mathscr A})}$ of the distribution of
${{\mathscr A}\!}$ can be obtained by setting  ${{\bf w} = m({\bf w})}$ and ${{\bf w} = \sigma({\bf w})}$:
\begin{equation}
\begin{array}{rclclcl}
m({\mathscr A})\!\!\!&=&\!\!\!f(m({\bf w}))\!\!\!&=&\!\!\!{\bf v}\, m({\bf w})\!\!\!&=&\!\!\text{a scalar} \\
\text{and}\;\;\;\sigma({\mathscr A})\!\!\!&=&\!\!\!\frac{\partial f}{\partial {\bf w}}\,\sigma({\bf w})\!\!\!&=&\!\!\!{\bf v}
\,\sigma({\bf w})\!\!\!&=&\!\!\text{a bivector}.
\end{array}
\end{equation}
The probability distribution of ${\mathscr A}$ is thus represented by the interval
\begin{equation}
\left[\,m({\mathscr A})-\sigma({\mathscr A}),\;m({\mathscr A})+\sigma({\mathscr A})\,\right].\label{int-2no}
\end{equation}
If we now assume that ${\bf w}$ is a unit bivector with a vanishing mean, then we have ${m({\mathscr A})=0}$
and ${\sigma({\mathscr A})={\bf v}}$, as in equation (\ref{asin23}) above.

\makeatletter 
\renewcommand*{\rightarrowfill@}{% 
  \arrowfill@\relbar\relbar\chemarrow} 
\makeatother 

It is instructive to note here that, geometrically, the propagation of error within ${S^3}$
is equivalent to a simple change in perspective:  
\begin{equation}
S^3\ni\,
  \underbrace{
    \overbrace{m({\bf w})}^\text{bivector} \;\pm\;\,
    \overbrace{\sigma({\bf w})}^\text{scalar}
   }_\text{quaternion}
\;\;\xrightarrow[]{\text{\;\;\;\;\;\;${f({\bf w})}$\;\;\;\;\;}}\;\;
\underbrace{
    \overbrace{m({\mathscr A})}^\text{scalar} \,\;\pm\;
    \overbrace{\sigma({\mathscr A})}^\text{bivector}
   }_\text{quaternion}.
\end{equation}
In particular, the probability distribution of ${\mathscr A}$ over ${S^3}$ corresponding to (\ref{int-2no}) is
equivalent to that of ${\bf w}$ over ${S^3}$ corresponding to (\ref{int-1no}).
\end{appendices}}

\end{document}